\documentclass[onecolumn,preprint,preprintnumbers,amsmath,amssymb]{revtex4}

\usepackage{graphicx}
\usepackage{dcolumn}
\usepackage{bm}

\begin{document}

\title{Tight focusing of polychromatic waves using angular spectrum compensation in spatially dispersive media}

\author{L.E. Helseth}

\affiliation{Department of Physics and Technology, University of Bergen, N-5007 Bergen, Norway}%

\begin{abstract}
A general method for tight focusing of waves, based on compensation of the angular spectrum, is established. We apply the method to monochromatic, polychromatic and diffusive waves. Diffusive and monochromatic waves may form spatially localized waves in free space, whereas polychromatic waves form light sheets or non-decaying traveling evanescent modes confined to subwavelength regions in media where the frequency depends on the wave vector. We suggest an analogy between our compensation method and the transformation of frequencies between inertial, relativistic coordinate systems.
\end{abstract}

\pacs{Valid PACS appear here}
\maketitle

\section{Introduction}
Physicists and engineers often need to find methods for confining a field as much as possible with a minimum of knowledge of the system at hand. In the case of monochromatic waves, it has been found that phase and amplitude wave engineering is of crucial importance for subwavelength focusing of waves\cite{Goodman,Pendry,Merlin}. In the case of pulsed waves, the situation is more complex, and optimal solutions have not been proposed to date, mainly due to the fact that the superposition of many spectral components in most cases tend to broaden the focal spot. Theoretically, one may hope to be able to focus strongly confined pulses as in Ref. \cite{Sherman}, but the aperture field and medium required remains elusive. However, there are a few promising pathways, e.g., using rainbow-colored apertures which combine to a single white-colored focus\cite{Zhu} or diffractive optical elements\cite{Yero}. Well developed theories and numerical algorithms are available for studying monochromatic and polychromatic waves in focal regions\cite{Sherman,Helseth1,Veetil,Bruegge}, but they do not provide direct insight into the reverse engineering problem and may not provide a fast solution or be optimal with the minimal amount of information given. Under such circumstances it is necessary to have straightforward methods which allows backtracking of the signal in a manner such that the initial field resulting in a strong focus is found. In this work we present a method to find the most strongly focused field. Here we extend the applicability of 'reverse' phase and amplitude engineering beyond previous approaches, and apply our method to monochromatic, polychromatic and diffusive waves.

\section{Diffusive waves}
\begin{figure}
\includegraphics[width=12cm]{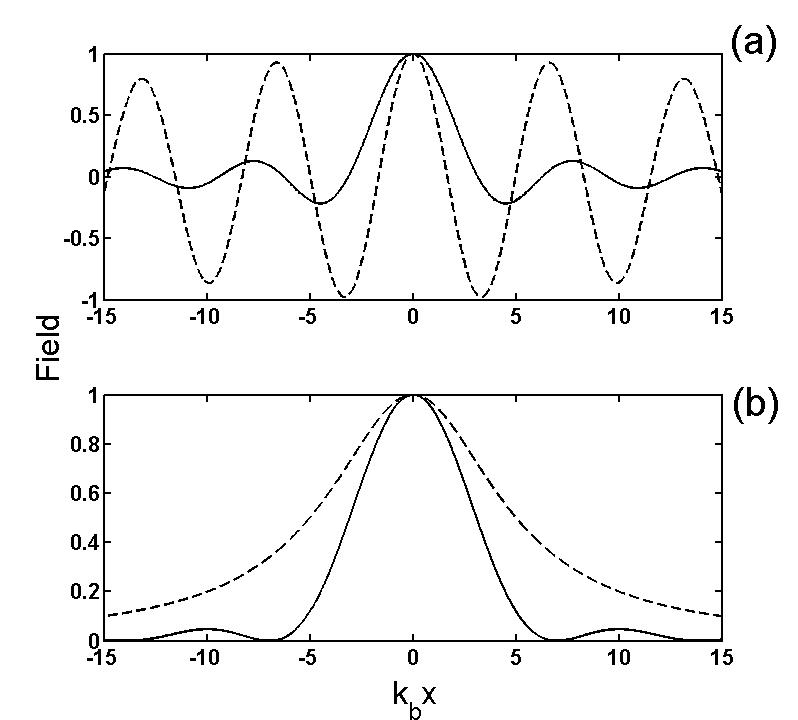}
\caption{\label{f1} In (a) the diffusive waves (a) with $k_0=1$ (a.u.), $D=1$ (a.u.) and $t_0=10$ are displayed.
In (b) the intensity of monochromatic, evanescent waves is displayed with $k_a=0.1$ (a.u.), $k_b=1$ (a.u.) and $k_bz_0 =5$ (a.u.). The solid lines show the field in focus, whereas the
dashed lines show the fields at $t=0$ for diffusive waves and $z=0$ for monochromatic evanescent waves.}
\vspace{2cm}
\end{figure}

We start by considering a related problem, namely the diffusion of a field in region of space containing no sources. The diffusion equation then describes the field according to
\begin{equation}
D\nabla ^{2} E(x,t)= \frac{\partial E (x,t)}{\partial t} \,\,\, .
\label{A}
\end{equation}
where D is the diffusion coefficient.
The field can now be written on the following general form
\begin{equation}
E (x,t) = \frac{1}{2\pi} \int_{-\infty}^{\infty}
E_{k}(k) e^{ikx -Dk^{2}t } dk  \,\,\, ,
\label{HHH}
\end{equation}
where $E_{k} (k)$ is the fourier amplitude. Usually, a solution of eq. \ref{HHH} will represent a diffusion process where the field delocalizes with increasing time. However, we are here interested in finding a field $E(x,0)$ with an optimally localized field $E(x,t_0)$ ($t_0>0$). To see how one may design such an initial field, let us assume that $E(k)=E_0\exp(Dk^{2}t_0)$ is a constant for $|k| \leq k_0$ and zero for $|k| \geq k_0$. Thus, the field is given by
\begin{equation}
E (x,t) = \frac{E_0}{2\pi} \int_{-k_0}^{k_0}
e^{ikx -Dk^{2}(t-t_0)} dk \,\,\, ,
\label{HH}
\end{equation}
The real value of eq. \ref{HH} at $t=0$ is given by
\begin{equation}
E (x,0) = \frac{E_0}{2\pi} \int_{-k_0}^{k_0}
cos(kx) e^{Dk^{2}t_0} dk\,\,\, ,
\label{HH}
\end{equation}
whereas at $t=t_0$ it is
\begin{equation}
E (x,t_0) =\frac{E_0}{2\pi} \int_{-k_0}^{k_0}
e^{ikx} dk =E_0\frac{sin(k_0 x)}{\pi x}\,\,\, ,
\label{H}
\end{equation}
By selecting $Dt_0$ such that $Dk_0^{2}t_0 \gg 1$, we see that the leading contribution to $E(x,0) \sim cos(k_0x) e^{Dk_0^{2}t_0}$, which is an oscillatory function with period $\Delta x= 2\pi/k_0$. Thus, the field at $t=0$ is strongly delocalized, but with sharp local fluctuations, as seen in Fig. \ref{f1} a). On the other hand, $E(x,t_0)$ is strongly localized with a central lobe $2\pi/k_0$, but is a factor $e^{Dk_0^{2}t_0}$ smaller than the field at $t=0$. A sharp localization is therefore obtained at the expense of an exponential reduction in field amplitude.

\section{Monochromatic waves}
The procedure introduced above can be straightforwardly applied to scalar monochromatic waves satisfying the Helmholtz equation
\begin{equation}
\left( \nabla ^{2} + k^{2} \right) E (x,z)=0 \,\,\, ,
\label{AA}
\end{equation}
where $k=\omega /v$ is the wavenumber, $\omega$ is the angular frequency and $v$ is the velocity of light in the medium under consideration.
Using the angular spectrum representation, the field can be written as
\begin{equation}
E (x,z) =\frac{1}{2\pi} \int_{-\infty}^{\infty}
E _{k} (k_{x})e^{ik_x x +ik_z z} dk_{x}\,\,\, ,
\label{BB}
\end{equation}
where the angular spectrum is given by
\begin{equation}
E_{k} (k_{x}) =\int_{-\infty}^{\infty} E (x',0)
e^{-ik_{x} x} dx' \,\,\, .
\label{CC}
\end{equation}
where
\begin{displaymath}
k_{z} = \left\{ \begin{array}{ll}
\sqrt{k^{2} -k_{x}^{2}} & \textrm{if $k^{2} \geq k_{x}^{2} $}\\
i\sqrt{k_{x}^{2} - k^{2}} & \textrm{if $k^{2} < k_{x}^{2} $}\\
\end{array} \right.
\label{DD}
\end{displaymath}
Here $k^{2} > k_{x}^{2}$ represent the homogenous plane
waves, whereas $k^{2} < k_{x}^{2}$ correspond to inhomogeneous or
evanescent plane waves that propagate in the $x$ direction and decay in the z - direction.

The trick is now to express the angular spectrum as
\begin{equation}
E_{k} (k_{x}) = E_{k}(k_x)e^{-ik_{z} z_0} \,\,\, ,
\label{perfect}
\end{equation}
where we must again require $E(k_x)$ to be bandlimited (i.e. nonzero only in a range $k_a \leq k_x \leq k_b$) and converges sufficiently quickly such that the solution for the field does not possess any unphysical divergency problem. This was discussed in Ref. \cite{Helseth}, and will not be repeated here. The field at the focal point ($z=z_0$) is now given by
\begin{equation}
E(x,z_0) =\frac{1}{2\pi} \int_{-\infty}^{\infty}
f _{k} (k_{x})e^{ik_x x} dk_{x}\,\,\, ,
\label{perfectfocus}
\end{equation}
i.e. a Fourier transform of the aperture function $f(k_x)$ resulting in a finite resolution at the focal point. As an example, assume now that $f_{k}(k_x)=1$ for $|k_x| \leq k_a$ ($k_a \ll k$) and zero elsewhere. Then we have, as in Eq. \ref{H},
\begin{equation}
E (x,z_0) =\frac{E_0}{2\pi} \int_{-k_0}^{k_0}
e^{ikx} dk =E_0\frac{sin(k_0 x)}{\pi x}\,\,\, ,
\label{H}
\end{equation}
In order to find the field $E(x,0)$ we must evaluate
\begin{equation}
E(x,0) =\frac{1}{2\pi} \int_{-k_0}^{k_0}
f_{k}(k_x) e^{ik_x x -ik_z z_0} dk_{x}\,\,\, .
\label{perfectfocus2}
\end{equation}
In Ref.\cite{Helseth} we found that in the case of paraxial waves ($k_x \ll k$) the field $E(x,0)$ is approximately a quadratic phase function, which has been described in detail in standard textbooks on optics\cite{Goodman}. One can also design focusing of evanescent waves using this approach. Let $f_k(k_x)=E_0$ if $k_a \leq k_x \leq k_b$ and zero elsewhere. Here $k_a \gg k$ and $E_0$ is a constant. The field can then be expressed as
\begin{equation}
E(x,z) \approx \frac{E_0}{2\pi}\left[ \frac{e^{ik_b x- k_b (z-z_0)} -e^{ik_a x- k_a (z-z_0)}}{ix -\left( z-z_0 \right)} \right] \,\,\, .
\label{perf}
\end{equation}
The intensity at the focus is therefore just $\propto sinc^{2}\left( \frac{k_b-k_a}{2}x \right) $ whereas the aperture field oscillates rapidly within an envelope such that the intensity is $\propto 1/(x^{2}+z_0^{2} )$. Thus, we have shown that the proposed method is in line with the theoretical observations made in e.g. Refs. \cite{Merlin,Helseth} for focusing of evanescent waves to subwavelength spots or lines. Notice that the aperture field is much stronger than the field in the focal region, i.e.  $E(x,0)/E(x,z_0) \sim e^{k_bz_0}$. If we assume that $k_b \gg k_a$, a possible criterium for focusing could be obtained by requiring that the aperture intensity (at $z=z_0$) envelope at $x=z_0$, corresponding to the position with half the maximum
intensity, should be wider than the focused intensity distribution of half-width $\sim \pi /k_b$. Thus, by requiring $k_b z_0 \geq \pi$, we ensure that the field is focused. At the same time, the field should not diverge at the aperture, and it is clear that $k_b z_0$ cannot be too large. Figure \ref{f1} b) shows the intensity $I \propto |E|^{2}$ at $z=z_0$ (solid line) and at $z=0$ (dashed line) when $k_b=1$, $k_a =0.1$ and $z_0=5$.

\section{Polychromatic waves}
An interesting question is whether the method above can be applied directly to polychromatic waves (e.g. pulsed waves).
To investigate this point, let us consider a time-dependent scalar field described by the wave equation as
\begin{equation}
\nabla ^{2} E (x,z,t)= \frac{1}{c^{2}} \frac{\partial ^{2} E(x,z,t)}{\partial t^{2}}
\label{timeA}
\end{equation}
In the angular spectrum representation, the field can be expressed as
\begin{equation}
E(x,y,z,t) = Re\left\{ \int_{0}^{\infty}
\int_{-\infty}^{\infty} E_{k} (k_{x}, \omega)
e^{ik_x x +ik_z z -i\omega (k_x) t} dk_{x}d\omega \right\} \,\,\, ,
\label{B}
\end{equation}
where $k=\omega(k_x)/c$ and
\begin{displaymath}
k_{z} = \left\{ \begin{array}{ll}
\sqrt{k^{2} -k_{x}^{2}} & \textrm{if $k^{2} \geq k_{x}^{2} $}\\
i\sqrt{k_{x}^{2} - k^{2}} & \textrm{if $k^{2} < k_{x}^{2} $}\\
\end{array} \right.
\label{C}
\end{displaymath}

In order to bring the polychromatic waves to focus we require that every spectral component is at focus a certain distance $z_1$ from the aperture at a given time $t_1$. This can be done by compensating the time and space parts of the phase separately, or by requiring a cross-compensation. The first alternative requires that we set
\begin{equation}
E_{k} (k_{x} ,\omega) = f(k_x,\omega)e^{-ik_{z} z_1 + i\omega t_1} \,\,\, ,
\label{phasecomp}
\end{equation}
which results in
\begin{equation}
E(x,y,z,t) = \int_{0}^{\infty}
\int_{-\infty}^{\infty} f (k_{x},\omega)
e^{ik_x x +ik_z (z-z_1) -i\omega (t-t_1)} dk_{x}d\omega  \,\,\, .
\label{phase2}
\end{equation}
It is seen that the field in focus $(z_1,t_1)$ is now just a sum over the frequencies of perfectly focused monochromatic wave components, which will be smeared out to a spot determined by the smallest wave vector of the spectrum (the largest wavelength in the case of homogeneous waves). In principle the procedure to find the aperture field is the same as that considered above for monochromatic waves. Unfortunately, it does not lend itself to immediate insight (except in the case of a rectangular spectrum for both $k_x$ and $\omega$, which results in similar fields as those given above), must in most cases be evaluated numerically, and will therefore not be pursued further here.

The second alternative is to require cross-compensation of the phase, which amounts to setting $k_z z_1 -\omega (k_x)t_1 =0$, which basically states that the spatial part of the phase is compensated by the temporal part at $(x,z_1,t_1)$. Such a requirement can only be fulfilled if $\omega$ depends on the spatial frequencies (i.e. direction of each plane wave), where the angular frequency is given by $\omega (k_x) =i\gamma (v_1) |k_x| v_1$. Here $\gamma (v_1) =1/\sqrt{1-(v_1/c)^{2}}$ and $v_1=z_1/t_1$.
The fact that frequency of the spatial wave vector depends on the direction was utilized in Ref. \cite{Zhu} to combine a rainbow spectrum to a single, focused spot. Conceptually the idea presented here is somewhat similar, although we use an entirely different approach to achieve the goal. Note that we must distinguish between the two cases $v_1 \geq c$ and $v_1 \leq c$.

Let us first consider the case $v_1 \geq c$. This can occur if the phase velocity of the wave exceeds the speed of light in vacuum. We then observe that $\omega (k_x)$ is real and given by $\omega (k_x)=\alpha k_x$ ($\alpha = v_1/\sqrt{(v_1 /c)^{2}-1}$). The scalar field is then given by
\begin{equation}
E(x,z,t) = \int_{0}^{\infty}
\int_{-\infty}^{\infty} E_{k} (k_{x},\omega)
e^{ik_x x + i|k_x| \alpha \left( t-\frac{z}{v_1} \right)} dk_{x}d\omega\,\,\, .
\label{E}
\end{equation}
It is seen that the fields are time and space-shifted such that $x_f=(\alpha /v_1)(z -v_1 t)$ gives transversal coordinate for the focal line. This suggests that the temporal field at $(t=0,x,z=0)$ can be imaged perfectly onto planes where $z>0$. Our theory has implicitly assumed, as is the case for all real systems, a bandlimited angular spectrum, e.g. nonzero in $k_a \leq k_x \leq k_b $. As an example, assume that $E_{k} (k_{x},\omega)=E_0$ (i.e. constant) in
$k_a \leq k_x \leq k_b $, such that the field is given by
\begin{equation}
E(x,z,t) \propto \frac{sin\left\{ \frac{k_b -k_a}{2} \left[ x+\alpha \left( t-\frac{z}{v_1} \right) \right] \right\}}{x+\alpha \left( t-\frac{z}{v_1} \right)} \,\,\, .
\label{F}
\end{equation}
We notice that due to the peculiar dispersion relationship required by the phase matching, the waves propagate as a light sheet located at the plane $z=(v_1/\alpha)(x+\alpha t)$. At the aperture, we require a pulse traveling in the negative x-direction. Oblique light sheets (in the case of electromagnetic waves) are then propagating outwards at an angle $v_1/\alpha$ with the aperture. It should be noted that upon imposing the requirement $\omega (k_x)=\alpha k_x$ without requiring additional boundary or media restrictions we obtain light sheets which somewhat unphysically extend to infinity (this problem could be removed by considering a more precise theory for causal and spatially dispersive media, but such an adventure is outside the scope of the current study). Only in the case $v_1\rightarrow c$ they are confined to the aperture plane at $t=0$, but this requires that a small change in the spatial frequency generates a very large change in frequency, thus putting significant constraints on the medium to be use.

A perhaps more interesting situation occurs when $v_1 \leq c$. Now $\alpha$ is purely imaginary, such that
\begin{equation}
E(x,z,t) = \int_{0}^{\infty}
\int_{-\infty}^{\infty} E_{k} (k_{x},\omega)
e^{ik_x x  -|k_x| \gamma \left( t-\frac{z}{v_1} \right)} dk_{x}d\omega \,\,\, .
\label{F1}
\end{equation}
As an example, we may approximate the angular spectrum as $E(k_x,\omega)=E_0$ for $k_a \leq k_x \leq k_b\, , \, \omega _1 \leq \omega \leq \omega _2$ and zero elsewhere. Here we also assume that
$k_a \gg k$, such that only evanescent waves are excited with $k_z\approx i|k_x|$ and $\omega (k_x) \approx i|k_x| v_1$. This evanescent-wave approximation therefore requires $v_1 \ll c$, and
the field will be
\begin{equation}
E(x,z,t) \propto  \left[ \frac{e^{ik_b x- k_b (z-v_1 t)} -e^{ik_a x- k_a (z-v_1 t)}}{ix -\left( z-v_1 t \right)} \right] \,\,\, .
\label{evantime}
\end{equation}
Note that eq. \ref{evantime} is identical to eq.\ref{perf} if we set $z_0=v_1 t$, and is therefore just a translation of the evanescent wave along the optical axis with the intensity distribution similar to that of fig. \ref{f1}. Since we require $k_b \gg k$,
which is fulfilled if we, as an example, set $k_b=10k=20\pi /\lambda$, the half-width of the intensity distribution at the traveling focal line
is narrower than $\sim \pi/k_b=\lambda /20$. When $t \gg 1/k_b v \sim t_1$ our approximate theory for evanescent waves does not longer hold in the vicinity of $z=0$ (this follows from the considerations above; see also Refs. \cite{Stamnes,Helseth} for a discussion about diverging solutions), where the intensity grows as $\exp(k_b v_1 t)/(x^{2} + (v_1t)^{2})$ with time. The fact that
the evanescent wave does not change at the focal line $z_1=v_1 t_1$ is surprising, given the condition $E(x,0,0)=E(x,z_1,t_1)$ above. However, we also note that $E(x,0,t_1) \gg E(x,z_1,t_1)$ such that the increasing energy of the field at the aperture is used to keep the field at the focal line $z_1=v_1 t_1$ unchanged in magnitude as it propagates outwards.

An interesting analogy occurs if one compares the problem of designing an aperture field that generate optimal focus with that of transformation of frequencies between inertial, relativistic coordinate systems\cite{Einstein}. To see this, consider a plane wave of the form $\exp(k_z z_1 -\omega (k_x)t_1 + \phi (k_x))$. Any wavepacket is a weighted sum over such plane waves in spatial and frequency coordinates. In order to bring the wave packet to focus we require that every spectral component is at focus a certain distance $z_1$ from the aperture at a given time $t_1$. This can be done by setting $k_z z_1 -\omega (k_x)t_1 + \phi (k_x)=0$, such that the phase is exactly compensated at $(z_1,t_1)$. It should be noted, as seen above, that such a requirement may give rise to strong localization at other positions as well. In any case, such a cross-compensation requirement can only be fulfilled if $\omega$ depends on the spatial frequencies (i.e. direction of each plane wave), where the frequency is given by
\begin{equation}
\omega (k_x) = -\frac{\phi (k_x)/t_1}{\left( \frac{v_1}{c}\right) ^{2} -1} \pm \frac{\phi (k_x)/t_1}{\left( \frac{v_1}{c}\right) ^{2} -1} \sqrt{1+\left[ \left( \frac{v_1}{c}\right) ^{2} -1 \right]
\left[ \left( \frac{k_x z_1}{ \phi (k_x)} \right) ^{2} -1 \right]} \,\,\, .
\label{timeC}
\end{equation}
Now consider the problem of transforming plane waves between inertial, relativistic coordinate systems. That is, consider an observer moving at a speed $v$ relative to a fixed frame. According to Einstein's special theory of relativity, the moving observer will detect a frequency $\omega '$ given by $\omega ' =\gamma (v) (\omega -k_z v)$, where $\gamma (v) =1/\sqrt{1-(v/c)^{2}}$\cite{Einstein}. From the above it may be inferred that this expression corresponds to $k_z z_1 -\omega (k_x)t_1 +i\phi (k_x) =0$ if we make the associations $\phi(k_x)/t_1 \rightarrow \omega'/\gamma (v_1)$ and $v_1 \rightarrow v$, i.e. the phase factor $\phi (k_x)$ must be associated with the frequency measured in the moving frame. In the special case considered in this study $\phi (k_x)=0$, which corresponds to $\omega ' =0$. For propagating waves we may set $k_z=(\omega/c)cos\theta$, where $\theta$ is the real angle at which a specific plane wave makes with the direction of motion. Then we must have $cos\theta =c/v$ and therefore $v\geq c$, in agreement with the observations made above.

\section{Conclusion}
In conclusion, we have suggested a phase-compensation method for designing aperture fields giving rise to strongly confined waves. The idea is to first try to compensate the phase or amplitude such that only transverse spatial frequencies are left in the angular spectrum representation at a given region in space, thus allowing a strongly focused wave to form here. Next we calculate the aperture field required to obtain such a phase or amplitude compensation. The method has been employed to study diffusive, monochromatic and polychromatic waves, and shown to give new insight into the problems at hand. The method here can probably also be applied to other wave systems where phase or amplitude compensation is beneficial.

\newpage

\newpage

\end{document}